\documentclass[floats,floatfix,showpacs,amssymb,prl,twocolumn,superscriptaddress,nofootinbib]{revtex4-1}
\usepackage[normalem]{ulem}
\usepackage{booktabs}
\usepackage{graphicx}%
\usepackage{dcolumn}%
\usepackage{bm}%
\usepackage{amsmath}
\usepackage{xcolor}
\usepackage{xspace}

\newcommand{\msun}{M$_\odot$\xspace}

\newcommand{\tjn}{\ensuremath{\theta_\mathrm{JN}}\xspace}

\newcommand{\va}{viewing angle\xspace}
\newcommand{\ia}{inclination angle\xspace}

\usepackage{hyperref}

\newcommand{\imr}{\texttt{IMRPhenomPv2}\xspace}

\newif\ifthetajn \thetajnfalse

\newcommand\prlsec[1]{\vspace{2mm}\noindent \textbf{\emph{#1}}--}

\begin{document}

\title{Systematic uncertainty of standard sirens from the viewing angle of binary neutron star inspirals}

\author{Hsin-Yu Chen}
\affiliation{Black Hole Initiative, Harvard University, Cambridge, Massachusetts 02138, USA}
\affiliation{LIGO Laboratory, Massachusetts Institute of Technology, Cambridge, Massachusetts 02139, USA}
\affiliation{Department of Physics and Kavli Institute for Astrophysics and Space Research, Massachusetts Institute of Technology, 77 Massachusetts Ave, Cambridge, MA 02139, USA}
\email{himjiu@mit.edu}
\thanks{NHFP Einstein Fellow}

\begin{abstract}
The independent measurement of Hubble constant with gravitational-wave standard sirens will potentially shed light on the tension between the local distance ladders and 
{Planck} experiments. Therefore, thorough understanding of the sources of systematic uncertainty for the standard siren method is crucial. 
In this paper, we focus on two scenarios that will potentially dominate the systematic uncertainty of standard sirens. 
First, simulations of electromagnetic counterparts of binary neutron star mergers suggest aspherical emissions, so the binaries available for the standard siren method can 
be selected by their viewing angles. This selection effect can lead to $\gtrsim 2\%$ bias in Hubble constant measurement even with mild selection. 
Second, if the binary viewing angles are constrained by the electromagnetic counterpart observations but the bias of the constraints is not controlled under $\sim 10^{\circ}$, 
the resulting systematic uncertainty in Hubble constant will be $>3\%$. In addition, we find that both of the systematics cannot be properly removed by the viewing angle 
measurement from gravitational-wave observations. Comparing to the known dominant systematic uncertainty for standard sirens, the 
$\leq 2\%$ gravitational-wave calibration uncertainty, the effects from viewing angle appear to be more significant. 
Therefore, the systematic uncertainty from viewing angle might be a major challenge before the standard sirens can resolve the tension in Hubble constant{, which is currently $\sim$9\%.}
\end{abstract}

\maketitle

\prlsec{Introduction}  
Gravitational-wave (GW) standard sirens provide an independent way to measure the Hubble constant ($H_0$), which is crucial for our understanding 
of the evolution of the Universe~\cite{1986Natur.323..310S,Abbott:2017xzu}. Currently, the $H_0$ 
measurements from cosmic microwave background~\cite{Aghanim:2018eyx} and some local distance ladders~\cite{2019ApJ...876...85R,2019arXiv190704869W,2020ApJ...891L...1P} appear to 
be inconsistent at $>2\sigma$ level. Independent $H_0$ measurement with the standard siren method has shown its potential 
to resolve the inconsistency~\cite{Abbott:2017xzu,2018Natur.562..545C}.

GW observations of compact binary mergers probe the luminosity distance ($D_L$) of the mergers directly. If the mergers also have 
electromagnetic (EM) counterparts~\cite{GBM:2017lvd}, e.g. short gamma-ray bursts (GRBs) or kilonova emissions that come with binary neutron star mergers (BNSs), 
the observation of the counterparts could allow for precise sky localization of the mergers and identification of the host galaxies~\cite{2005ApJ...629...15H,2006PhRvD..74f3006D}. 
With the luminosity distance of the GW source and the redshift of the host galaxy, cosmological parameters can be constrained. 
This is the so-called standard siren method with the use of EM counterparts.
GW170817 was the first successful standard siren~\cite{Abbott:2017xzu}. {Several forecasts predict that a $2\%$ $H_0$ measurement can be achieved by combining 
$\sim$50 BNSs with identified host}~\cite{2013arXiv1307.2638N,2018Natur.562..545C,2019PhRvL.122f1105F}. 

In order to resolve the $H_0$ controversy, the systematic uncertainty in the standard siren method has to be well-understood. 
One dominant systematics comes from the calibration of amplitude measurement of GW signals. The calibration uncertainty 
currently leads to $\leq 2\%$ systematics in the GW distance measurement, while this uncertainty is expected to reduce in the future~\cite{2016RScI...87k4503K,2020arXiv200502531S}. 
Another source of systematics comes from the reconstruction of the peculiar velocity fields around the host galaxies~\cite{2020MNRAS.492.3803H,2019arXiv190908627M,2020MNRAS.tmp.1270N}. 
{This systematic is remarkable for nearby events, while the majority of events are expected to lie at further distances and less affected by the uncertainty of peculiar motions.}
Other known sources of systematic uncertainty, e.g. the accuracy of GW waveforms~\cite{Abbott:2018wiz}, are expected 
to play a secondary role.
      
In this paper, we highlight two sources of systematic uncertainties for standard sirens that have not been thoroughly discussed before. 
Both of the systematics are related to the EM counterpart observations and the \va of the binaries ($\zeta$)
~\footnote{Since the EM counterpart emissions barely depend on the direction of the binary rotation (clockwise or counterclockwise), 
in this paper we define the \va as $\zeta\equiv \mathrm{min}(\tjn, 180^\circ - \tjn)$, where $\tjn$ denotes the \ia of the binary.}:
First, simulations of BNSs suggest that their EM emissions are likely aspherical~\cite{2011ApJ...736L..21R,2017LRR....20....3M,2019MNRAS.489.5037B,2020arXiv200200299D}. 
For example, the brightness of kilonovae can have a factor of 2-3 angular dependent variation. The color of kilonovae can also change with the \va. 
{Therefore the EM-observing probability for BNSs can depend on the \va (e.g.,~\cite{2020arXiv200406137S}).} 
If this \va selection effect is not accounted for correctly, 
$H_0$ measurement will 
be biased after combining multiple standard sirens. Second, EM observations of BNSs provide constraint on the \va. 
The \va of BNS GW170817~\cite{TheLIGOScientific:2017qsa} has been reconstructed from the profiles of its EM emissions~\cite{2018ApJ...860L...2F,2020ApJ...888...67D} and from the observations of 
the jet motions~\cite{2018Natur.561..355M}. These reconstructions help breaking the degeneracy between the luminosity distance and \ia of BNSs in 
GW parameter estimations~\cite{2019PhRvX...9c1028C}, improving the 
precision of distance measurement, and reducing the $H_0$ measurement uncertainty~\cite{2017ApJ...851L..36G,2019NatAs...3..940H}. 
However, if the EM constraints on the \va are systematically biased, the distance and $H_0$ estimation will also be biased. 

We find that both of the systematics can yield significant bias in $H_0$ measurement, undermining the standard siren's potential 
to resolve the $H_0$ tension. Since both of the scenarios we discuss originate from the uncertainty of EM observations, we also explore if it is possible to independently determine 
the systematics by analyzing the {\va measured from GWs}. 
Unfortunately, most of the events suffer from the large uncertainty of the estimations and the systematics can be 
difficult to disclose.

\prlsec{Simulations} We simulate 1.4\msun-1.4\msun non-spinning BNS detections with the \imr waveform and assumed a network signal-to-noise ratio of 12 {GW-detection} threshold. 
We use Advanced LIGO-Virgo O4 sensitivity~\cite{Aasi:2013wya} for the simulations~\footnote{Specifically, 
the \texttt{aligo\_O4high.txt} file for LIGO-Livingston/LIGO-Hanford, and \texttt{avirgo\_O4high\_NEW.txt} for Virgo in this document: \url{https://dcc.ligo.org/LIGO-T2000012/public}.}. 
With this sensitivity, it is valid to assume the BNS astrophysical rate does not evolve over redshifts, and the BNSs are uniformly distributed in comoving volume before detections. 
Planck cosmology is used ($H_0=67.4$ km/s/Mpc, $\Omega_m=0.315$, $\Omega_k=0$)~\cite{Aghanim:2018eyx}. 
Suppose the data from GW and EM are denoted as $\mathcal{D}_{\rm GW}$ and $\mathcal{D}_{\rm EM}$ respectively,
one can follow ~\cite{2018Natur.562..545C,2019MNRAS.486.1086M} to write down the $H_0$ likelihood for single event as:
\begin{equation}\label{eq:likelihood}
p(\mathcal{D}_{\rm GW},\mathcal{D}_{\rm EM}|H_0)=\frac{\displaystyle\int p(\mathcal{D}_{\rm GW}|\vec{\Theta})p(\mathcal{D}_{\rm EM}|\vec{\Theta})p_{\rm pop}(\vec{\Theta}|H_0)d\vec{\Theta}}{\displaystyle\int p_{\rm det}(\vec{\Theta})p_{\rm pop}(\vec{\Theta}|H_0)d\vec{\Theta}},
\end{equation}
where $\vec{\Theta}$ represents all the binary parameters, such as the mass, spin, luminosity distance, sky location, and \ia etc.. 
$p_{\rm pop}(\vec{\Theta}|H_0)$ is proportional to the abundance of binaries with parameters $\vec{\Theta}$ in the Universe. 
\begin{equation}\label{eq:det}
p_{\rm det}(\vec{\Theta})\equiv \displaystyle \iint\limits_{\substack{{\hat{\mathcal{D}_{\rm GW}}>{\rm GW}_{\rm th}}, \\{\hat{\mathcal{D}_{\rm EM}}>{\rm EM}_{\rm th}}}} p(\hat{\mathcal{D}_{\rm GW}}|\vec{\Theta})p(\hat{\mathcal{D}_{\rm EM}}|\vec{\Theta})d\hat{\mathcal{D}_{\rm GW}}d\hat{\mathcal{D}_{\rm EM}}, 
\end{equation}  
in which the integration only goes over data above the {GW- and EM-}detection threshold, ${\rm GW}_{\rm th}$ and ${\rm EM}_{\rm th}$. 
{We note that $\mathcal{D}_{\rm GW}$ and $\mathcal{D}_{\rm EM}$ in the numerator of Equation~\ref{eq:likelihood} are the data from the detections, 
so they are above the detection threshold by definition.}  

For GW likelihood $p(\mathcal{D}_{\rm GW}|\vec{\Theta})$ the relevant binary parameters are the luminosity distance ($D_L$) and the \ia (\tjn), 
so we use the algorithms developed in ~\cite{2019PhRvX...9c1028C} to estimate $p(\mathcal{D}_{\rm GW}|D_L,\tjn)$
~\footnote{Note that this step also generates the distance-\ia posterior $p(D_L,\tjn|\mathcal{D}_{\rm GW})$ when the likelihood is multiplied by a prior, which will be 
used in later part of this paper.}.  
We will discuss the EM likelihood $p(\mathcal{D}_{\rm EM}|\vec{\Theta})$ and how it affects the $H_0$ estimation in the next two sections. 
We use the $H_0$ posterior of GW170817~\cite{Abbott:2017xzu} as the prior and combine multiple $H_0$ likelihoods from simulated 
events to produce the final $H_0$ posterior. We repeat the simulations 100 times and report the average for the results throughout this paper.

\prlsec{Systematics from {viewing angle} selection effect} 
If the EM counterpart emissions are aspherical, BNSs with some viewing angles could be easier to observe than from other directions. 
How the {EM-observing} probability depends on the \va should be included in the EM likelihood $p(\mathcal{D}_{\rm EM}|\vec{\Theta})$. 
However, if such dependency is unknown or ignored, Equation~\ref{eq:likelihood} and the combined $H_0$ posteriors from multiple events will be incorrect.

\begin{figure}
\centering
\includegraphics[width=1.0\columnwidth]{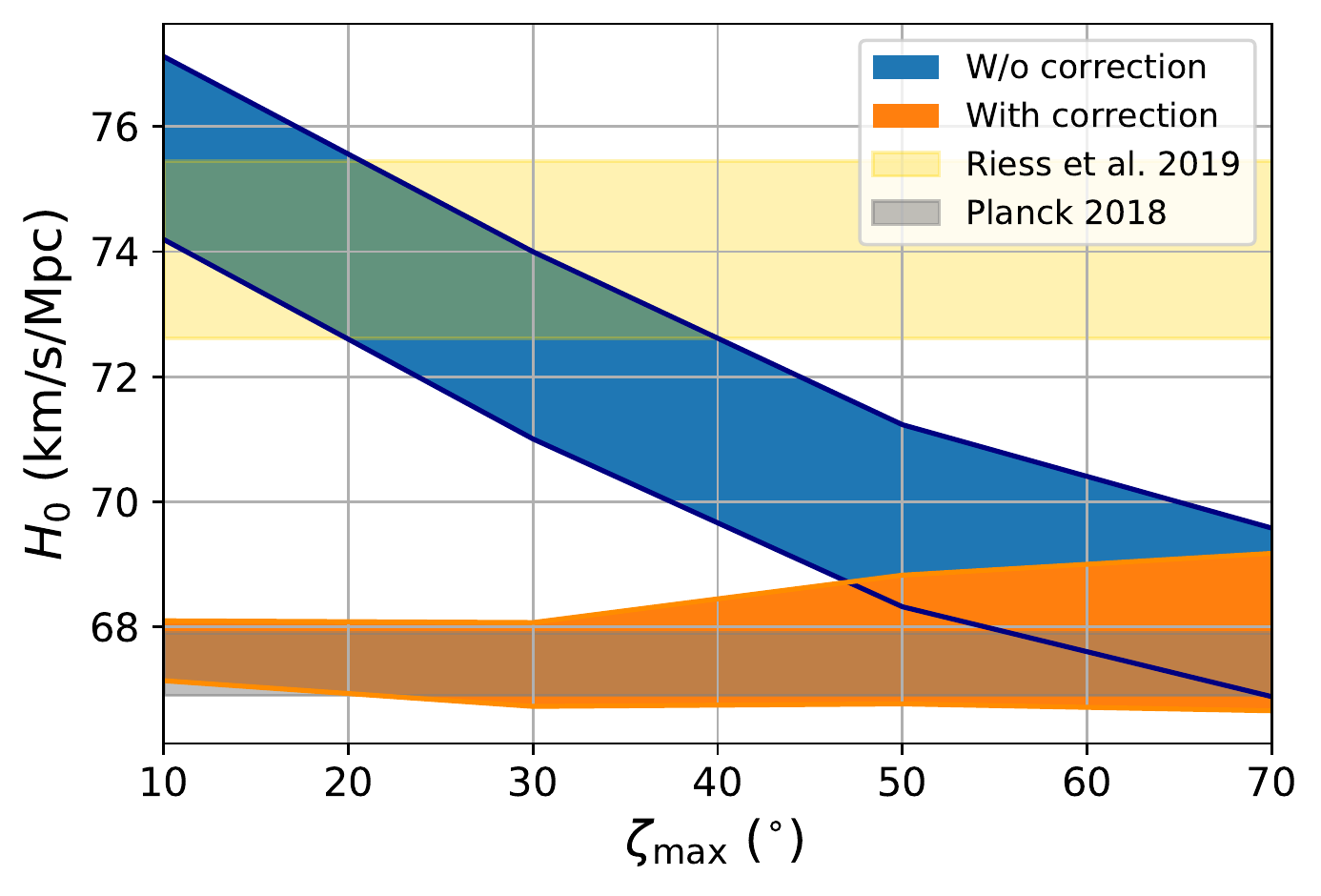}
\caption{\label{fig:selection}
Hubble constant measurement uncertainty (1-$\sigma$) from 50 standard sirens as a function of the maximum viewing angle of the binaries. 
The Hubble constant used for the simulations is $67.4$ km/s/Mpc. 
If the maximum viewing angle is known, appropriate corrections can be applied (as described in Equation~\ref{eq:likelihood}) and the uncertainty is the 
\emph{With correction} band. In contrast, the \emph{W/o correction} band shows the level of bias if the maximum viewing angle is unknown. 
For reference, the two horizontal 
bands denote the $H_0$ reported by Riess et al.~\cite{2019ApJ...876...85R} ($74.03\pm 1.42$ km/s/Mpc) and 
Planck~\cite{Aghanim:2018eyx} ($67.4 \pm 0.5$ km/s/Mpc).
}
\end{figure}

{How the EM-observing probability depending on the \va is determined by the EM emissions, the EM facilities and the observing strategies. 
Here we explore two generic examples}
~\footnote{Since a telescope, an EM model, and an EM serach pipeline have to be specified before the noise properties of EM data 
can be quantified, in this paper we assume there is no EM observing noise for simplification.}: {In the first example}, we assume 
only BNSs with \va less than $\zeta_{\rm max}$ are observable in EM. Smaller $\zeta_{\rm max}$ represents stronger selection since the \va is more limited. 
{Short GRBs with beamed emissions are likely to lead to such abrupt decay in EM-observing probability beyond the beaming angle.}
In Figure~\ref{fig:selection} we show the symmetric 1-$\sigma$ uncertainty in $H_0$ for different $\zeta_{\rm max}$ if 50 events are combined. 
{If this selection on \va is unknown or ignored}, we find the $H_0$ measurement significantly biased even if $\zeta_{\rm max}$ is as 
large as $\sim 60^{\circ}$ (the band \emph{W/o correction}). Only as a demonstration, we also show the $H_0$ uncertainty assuming 
the \va selection $\zeta_{\rm max}$ is perfectly known (the band \emph{With correction}). 
If $\zeta_{\rm max}$ is known, $p(\mathcal{D}_{\rm EM}|\vec{\Theta})$ is taken as 0 when $\zeta > \zeta_{\rm max}$.

In the second example, we assume the {EM-observing} probability is a continuous function of \va and the EM likelihood is 
taken as $p(\mathcal{D}_{\rm EM}|\vec{\Theta})=0.5 ({\rm cos}(\zeta)+1)$. With this assumption, all face-on binaries are observable, while 
only 50\% of edge-on binaries can be observed.
{Aspherical kilonova emission can result in this continuous observing function (e.g., ~\cite{2020arXiv200406137S}).}
 Without correction, we find the 1-$\sigma$ uncertainty in $H_0$ for 50 events 
lying between $[67.5,70.2]$km/s/Mpc, equivalent to $\sim 2\%$ bias in $H_0$. 

\begin{figure}
\centering
\includegraphics[width=1.0\columnwidth]{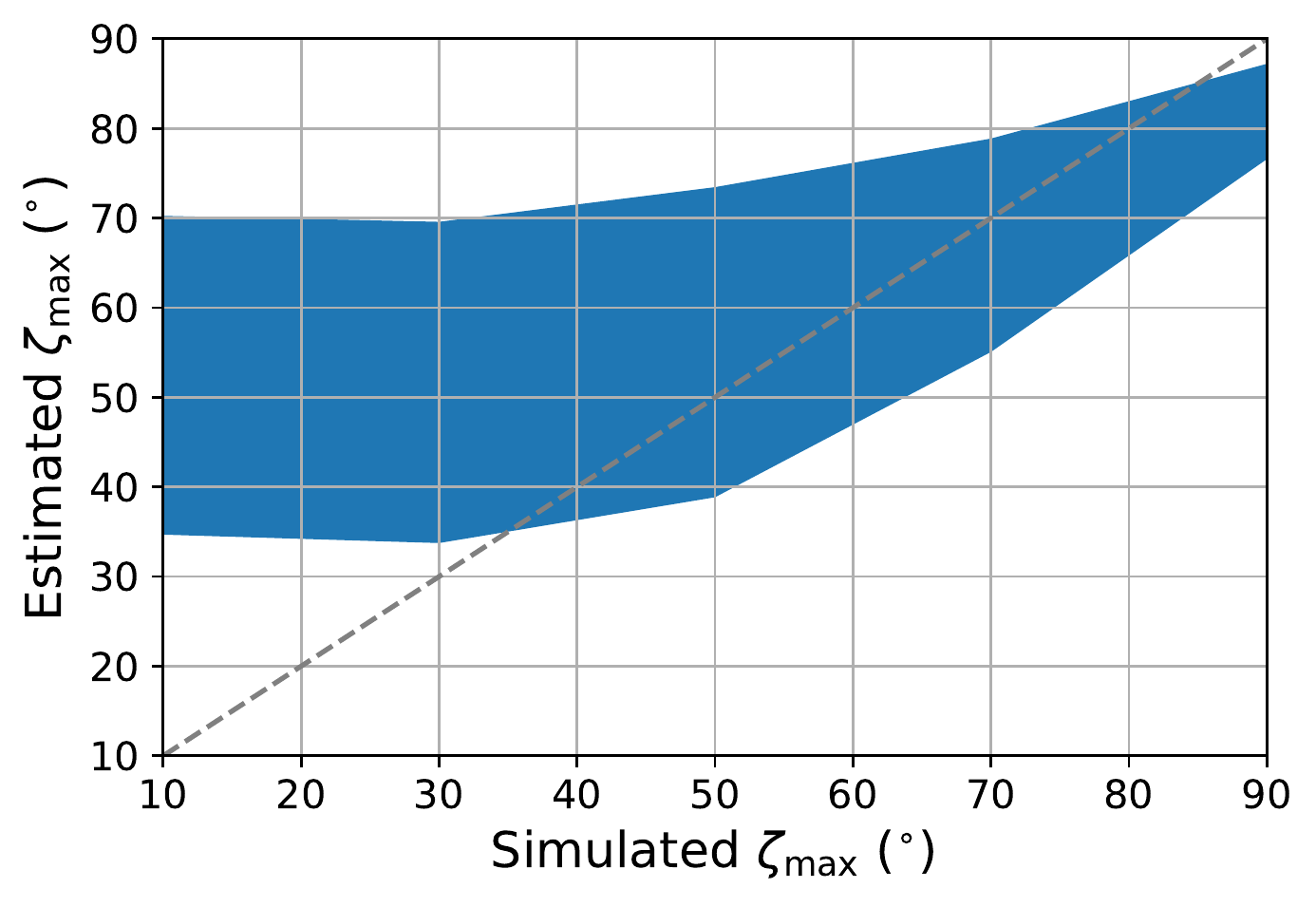}
\caption{\label{fig:selectionmeasure}
Maximum \va $\zeta_{\rm max}$ estimated from 50 BNSs' {GW-\va} posteriors. The band denotes the symmetric $1\sigma$ uncertainty of the 
estimations, and the grey dashed line is the equal-axis line to guide the eye. 
Small simulated $\zeta_{\rm max}$ 
are not estimated accurately due to large uncertainty of the \va posteriors. 
}
\end{figure}

A possible way to determine the \va selection effect is to analyze the {\va measurements from GWs for events with EM counterparts.}
We try to estimate $\zeta_{\rm max}$ in the first example above from the GW data of $N$ events \{${\mathcal{D}_1,\mathcal{D}_2...\mathcal{D}_N}$\}:
\begin{align}\label{eq:selectionmeasure}
\nonumber &p(\zeta_{\rm max}|\mathcal{D}_1,\mathcal{D}_2...\mathcal{D}_N)=\frac{p(\zeta_{\rm max})\displaystyle \prod_{k=1}^{N} p(\mathcal{D}_k|\zeta_{\rm max})}{\displaystyle\prod_{k=1}^{N} p(\mathcal{D}_k)} \\
\nonumber &=p(\zeta_{\rm max})\displaystyle\prod_{k=1}^{N}\displaystyle \int_0^{\pi/2}\frac{p(\zeta|\mathcal{D}_k)p(\zeta_{\rm max}|\zeta,\mathcal{D}_k)}{p(\zeta_{\rm max})}d\zeta  \\
\nonumber &=p(\zeta_{\rm max})\displaystyle\prod_{k=1}^{N}\displaystyle \int_0^{\pi/2}\frac{p(\zeta|\mathcal{D}_k)p(\zeta|\zeta_{\rm max})}{p(\zeta)}d\zeta  \\
&=p(\zeta_{\rm max})\displaystyle\prod_{k=1}^{N}\frac{\displaystyle \int_0^{\zeta_{\rm max}}p(\zeta|\mathcal{D}_k)d\zeta}{\displaystyle \int_0^{\zeta_{\rm max}} p(\zeta) d\zeta}.
\end{align} 
The first line comes from the fact that each event are independent. The third line considers $p(\zeta_{\rm max}|\zeta,\mathcal{D}_k)=p(\zeta_{\rm max}|\zeta)$, 
and the last line takes $p(\zeta|\zeta_{\rm max})\propto p(\zeta)$ for $\zeta<\zeta_{\rm max}$. Equation~\ref{eq:selectionmeasure} 
can then be calculated from the {GW-\va} posterior $p(\zeta|\mathcal{D}_k)$, which is obtained by integrating 
the distance-inclination angle posterior $p(D_L,\tjn|\mathcal{D}_k)$ over $D_L$, and the prior on \va $p(\zeta)$~\cite{2011CQGra..28l5023S}. 
Without any prior on $\zeta_{\rm max}$, i.e. $p(\zeta_{\rm max})$ is taken as a constant, 
in Figure~\ref{fig:selectionmeasure} we show the symmetric 1-$\sigma$ uncertainty of the $\zeta_{\rm max}$ posteriors (Equation~\ref{eq:selectionmeasure}) as a function of the maximum EM \va of 50 
simulated BNSs.  
We find that $\zeta_{\rm max}$ can only be confined to $\sim 20^{\circ}$ 1-$\sigma$ uncertainty. In addition, 
the estimated $\zeta_{\rm max}$ is biased for small $\zeta_{\rm max}$ because {GW-\va} posteriors typically peak around $30^{\circ}$ 
with about $20^{\circ}$ uncertainty~\cite{2019PhRvX...9c1028C}. Small $\zeta_{\rm max}$ is therefore difficult to reconstruct even if all BNSs 
with observable EM counterparts are face-on/off.

\prlsec{Systematics from biased {EM-constraint} on viewing angle} 
{Another possible bias comes from the interpretation of the EM observations. }The angular dependency of EM emissions can be used to estimate the \va of BNSs. 
However, lack of robust understanding of the EM emission model can lead to biased interpretation of the \va. 

\begin{figure}
\centering
\includegraphics[width=1.0\columnwidth]{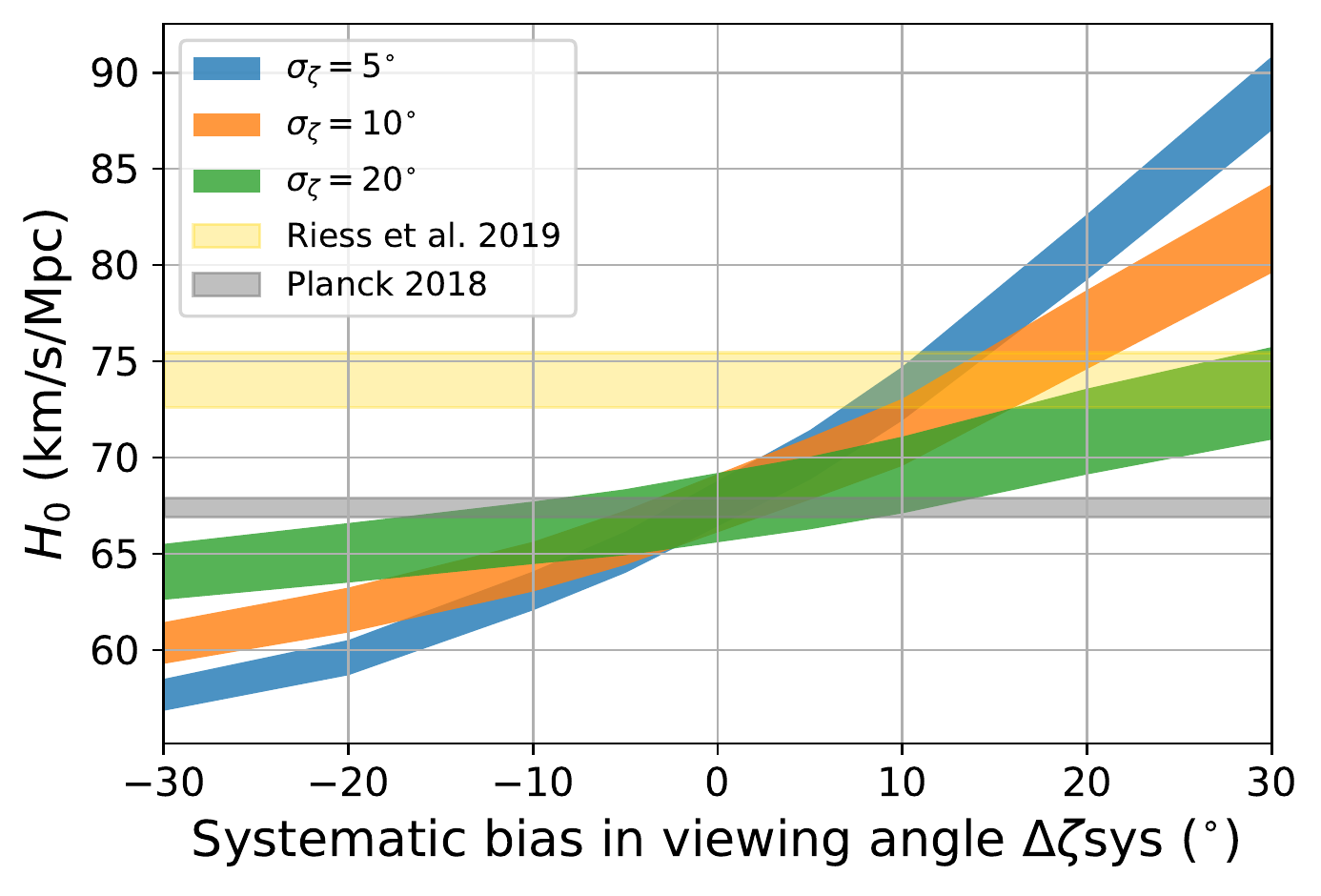}
\caption{\label{fig:extbias}
Hubble constant measurement uncertainty (1-$\sigma$) from 20 standard sirens as a function of the systematic bias in the 
binary viewing angle constrained by EM observations. 
Three different statistical uncertainties in the EM-constrained viewing angle ($\sigma_{\zeta}=5^{\circ},10^{\circ},20^{\circ}$) are shown.
The Hubble constant used for the simulations is $67.4$ km/s/Mpc. 
}
\end{figure}

Suppose the EM observations suggest a \va of $\zeta_{\rm EM}$ with 1-$\sigma$ uncertainty 
of $\sigma_{\zeta}$, the EM likelihood in Equation~\ref{eq:likelihood} is then proportional to
\begin{equation}\label{eq:emconstraint}
 p(\mathcal{D}_{\rm EM}|\vec{\Theta})\propto \begin{cases}
\cal{N}(\tjn;\zeta_{\rm EM},\sigma_{\zeta}) &\text{if}\, 0\leq \tjn \leq \pi/2 \\
\cal{N}(\tjn;\pi-\zeta_{\rm EM},\sigma_{\zeta}) &\text{if}\, \pi/2< \tjn \leq \pi,
\end{cases}
\end{equation}
where $\cal{N}(\tjn;\zeta_{\rm EM},\sigma_{\zeta})$ denotes a normal distribution with mean $\zeta_{\rm EM}$ and standard deviation $\sigma_{\zeta}$ evaluated at $\tjn$. 
Since Equation~\ref{eq:emconstraint} provides constraint on the inclination angle and reduces the binary parameter space in Equation~\ref{eq:likelihood}, 
the Hubble constant can be measured more precisely~\cite{2019PhRvX...9c1028C}. 
However, if the EM constraint on the \va is off by 
\begin{equation*}
\Delta \zeta_{\rm sys}\equiv \zeta_{\rm EM}-\zeta_{\rm real},
\end{equation*}
 where $\zeta_{\rm real}$ denotes the real \va of the event, the $H_0$ measurements will be biased. 
 For single event the bias in $H_0$ may not be obvious, because the statistical uncertainty in $H_0$ 
 dominates the overall uncertainty. The bias will become clear after the $H_0$ posteriors are combined over multiple events. 
 In Figure~\ref{fig:extbias} we show the extent of overall bias in $H_0$ if the EM constraint on \va is always off by $\Delta \zeta_{\rm sys}$ for 20 events.

When the viewing angles are overestimated (underestimated), the combined $H_0$ is overestimated 
(underestimated). Smaller $\sigma_{\zeta}$ affects the $H_0$ measurement more significantly 
for the same $\Delta \zeta_{\rm sys}$. Note 
that $\Delta \zeta_{\rm sys}$ is not necessarily a constant across different events. We choose a 
fixed bias $\Delta \zeta_{\rm sys}$ across 20 events only to reveal the \emph{average} level of $H_0$ bias 
for a given $\Delta \zeta_{\rm sys}$, since the systematic uncertainty in $H_0$ is not expected to evolve with the total number of events. 
Combining 20 events instead of using a single event reduces the statistical fluctuations and manifest the systematic uncertainty 
in $H_0$. 
Our simulations map the systematic uncertainty in \va {inferred from EM observations} to $H_0$ for the first time. 
From Figure~\ref{fig:extbias}, $\Delta \zeta_{\rm sys}$ has to be $\lesssim 10^{\circ}$ to be accurate enough to address the tension between \emph{Planck} and the local distance ladders. 

\begin{figure}
\centering
\includegraphics[width=1.0\columnwidth]{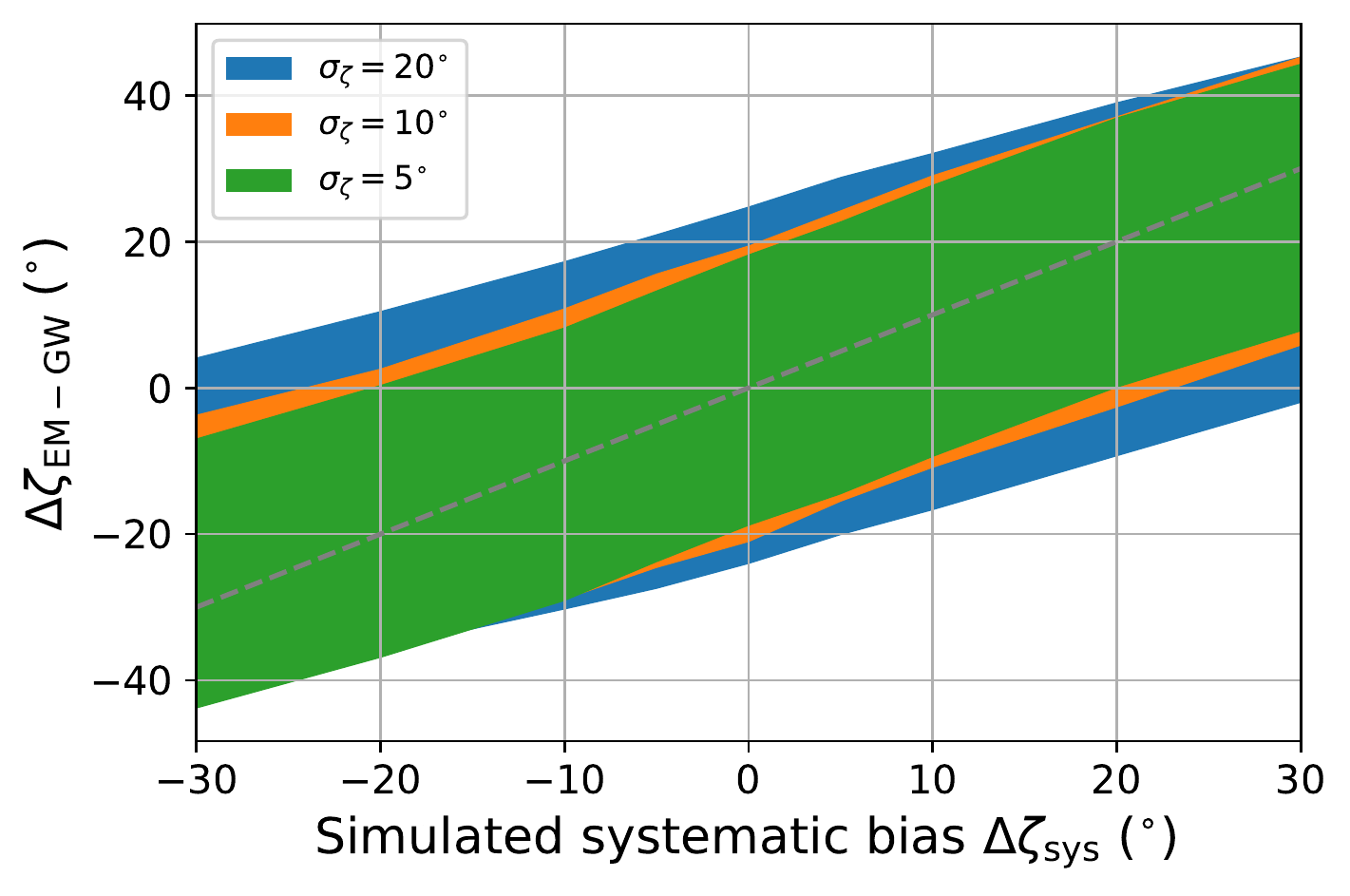}
\caption{\label{fig:biasmeasure}
The average difference between {EM- and GW-\va} posteriors $\Delta \zeta_{\rm EM-GW}$ for 20 BNSs with EM posteriors systematically off by $\Delta \zeta_{\rm sys}$. 
The $1-\sigma$ uncertainty of the difference for three EM posterior statistical uncertainties, $\sigma_{\zeta}=5^{\circ},10^{\circ},20^{\circ}$, are $18.5^{\circ}$, $20^{\circ}$, and $24^{\circ}$, respectively.  
The grey dashed line is the equal-axis line to guide the eye.
}
\end{figure}

Next, we wonder if a comparison between the GW and EM measurement of the \va will help disclosing the bias in EM interpretations. 
Suppose the \va posteriors from GW and EM for a BNS are $\Upsilon(\zeta)$ and $\varepsilon(\zeta)$ respectively, we can define their difference as 
\begin{equation}\label{eq:diff}
\Delta \zeta_{\rm EM-GW}\equiv \displaystyle \int_0^{\pi/2} \int_0^{\pi/2} (\zeta_2-\zeta_1)\times \Upsilon(\zeta_1)\times \varepsilon(\zeta_2)\, d\zeta_1 d\zeta_2 .
\end{equation}
We find that the average of $\Delta \zeta_{\rm EM-GW}$ over 20 BNSs traces $\Delta \zeta_{\rm sys}$ with $1-\sigma$ 
uncertainty $>18^{\circ}$, as shown in Figure~\ref{fig:biasmeasure}. This uncertainty of $\Delta \zeta_{\rm EM-GW}$ is larger than the required 
accuracy of $\zeta_{\rm EM}$ above, making it difficult to resolve the $H_0$ systematics from biased EM constraint.

\prlsec{Discussion}
In this paper we evaluate the extent of bias in $H_0$ as a result of the geometry of EM emissions from BNSs. 

{If the geometry affects the EM-observing probability, the selection effect can introduce a $\gtrsim 2\%$ bias on $H_0$.}
The example of maximum viewing angle we present may happen due to the choice of  
kilonova observing strategies or the sharp decline beyond a viewing angle for short GRB emission. 
Future studies of the jet structure of GRBs will be crucial to correct the selection effect for standard sirens.
On the other hand, the example of continuous \va selection is relevant for kilonova observations.  
Simulations show that edge-on BNSs are more difficult to localize~\cite{2019PhRvX...9c1028C}, and their kilonova emissions can be redder and dimmer~\cite{2020arXiv200200299D}. 
The resulting selection effects will depend on the telescopes, the observing strategies, and the observing conditions, so the overall effects 
can be subtle to estimate {and correct}. 

{Even if the selection effect is corrected, when} the geometry of EM emissions is used to confine the BNSs' \va, the systematic uncertainty in \va introduced by the EM 
interpretations has to be less than $\sim 10^{\circ}$. Since the binary rotational axis doesn't have to be perfectly aligned with the 
major axis of EM emissions, and the geometry of EM emissions is unknown, to control the systematics of EM inferred \va can 
be challenging. We also show that the comparison between {EM- and GW-\va measurements} can help estimating the systematics, but the 
precision of the estimation may not be good enough to completely remove the bias.

We note that in reality other binary parameters will also affect the {EM-observing} probability. 
Therefore, more complete considerations of EM models and projections of {EM-observing} probability for future telescopes involved in the search for EM counterparts 
will result in more accurate estimation of the bias in $H_0$. 
Unlike the \va measurement, some parameters, such as the mass, are estimated precise enough from GW signals for the selection effect to be taken care of. 
Overall, we find the selection over \va discussed in this paper the most subtle and difficult to resolve.

If the \va selection effect is significant, it is possible to reconstruct the selection by comparing the number of BNSs with and without 
EM counterparts. The distribution of \va for BNSs detected in GWs is well-understood~\cite{2011CQGra..28l5023S}. For example, 
it is known that about 15\% of BNS detections have \va larger than $60^{\circ}$. If 15\% of BNSs miss counterparts, one explanation is that the maximum EM \va is 
around $60^{\circ}$. A reconstruction of short GRB viewing angles {using the inclination angles and distances of GW-GRB joint detections} has been shown in ~\cite{2019arXiv191204906F}. 
However, the reconstruction for kilonova will be more difficult since their {EM-observing} probability has more complicated dependency on 
the \va. Such reconstruction can also be easily contaminated by other factors 
that affect the {EM-observing} probability and will have to be evaluated carefully. 

Although our discussion focuses on BNSs, there are simulations suggesting stronger \va dependency 
for EM counterparts of neutron star-black hole mergers~\cite{2020arXiv200200299D}. Therefore neutron 
star-black hole mergers can possibly introduce larger bias when they are used as standard sirens~\cite{2018PhRvL.121b1303V}.

We note that the standard siren method we discuss in this paper relies on the observations of EM counterparts and the measurements 
of the BNSs' redshift. A complimentary approach of the standard siren method doesn't require the EM counterparts but make use of 
galaxy catalogs may help deducing the systematics discussed in this paper. However, the {galaxy catalogs} 
approach will suffer from lower $H_0$ precision and other sources of systematics~\cite{2018Natur.562..545C,2019arXiv190806050G}, 
making it complicated to contribute to the issues.  

Finally, the calibration uncertainty in GWs currently dominates the known systematic uncertainty for standard sirens. The bias in $H_0$ 
from calibration can be as large as $\sim 2\%$~\cite{2016RScI...87k4503K,2020arXiv200502531S}. Both of the systematics we find in this work can introduce $H_0$ bias larger than 2\%. 
In summary, the systematic uncertainty from \va for standard sirens can be a major challenge to resolve the tension in Hubble constant, and 
we look forward to future development to explore this topic.

\prlsec{acknowledgments} We acknowledge valuable discussions with Sylvia Biscoveanu, Michael Coughlin, Carl-Johan Haster, Daniel Holz, Kwan-Yeung Ken Ng, and  Salvatore Vitale.
HYC was supported by by the Black Hole Initiative at Harvard University, which is funded by grants from the John Templeton Foundation and the Gordon and Betty Moore Foundation to Harvard University.

\bibliography{references}

\end{document}